\documentstyle[12pt,epsfig]{article}

\def\ra{\rightarrow}

\def\be{\begin{equation}}
\def\ee{\end{equation}}
\def\bea{\begin{eqnarray}}
\def\eea{\end{eqnarray}}

\def\ra{\rightarrow}

\def\bea{\begin{eqnarray*}}
\def\ena{\end{eqnarray*}}
\def\beq{\begin{equation*}}
\def\enq{\end{equation*}}

\def\TT{\textstyle}
\def\D0{D\O~}

\begin{document}

\thispagestyle{empty}
\setcounter{footnote}{1}
\renewcommand{\thefootnote}{\fnsymbol{footnote}}

\def\today{\number\day
    \space\ifcase\month\or
      January\or February\or March\or April\or May\or June\or
      July\or August\or September\or October\or November\or
             December\fi
    \space\number\year}
\def\thisday{April 5, 1999}

\begin{flushright}
{\noindent
{hep-ph/9905391} \hfill {CTEQ-904}\\
{April 1999} \hfill {MSUHEP-90414}
}
\end{flushright}

\vspace{0.5cm}

\begin{center}
{ {\large {\bf
New Fits for the Non-Perturbative Parameters in the  \\[0.18cm]
CSS Resummation Formalism}}} \\[1.2cm]

\begin{center}
{ {\sc F. Landry}, {\sc R. Brock}, {\sc G. Ladinsky},
and  {\sc C.-P. Yuan} }
\end{center}

\begin{center}
{\it Department of Physics and Astronomy,
Michigan State University, \\
East Lansing, MI 48824, USA}
\end{center}

\vspace{0.4cm}
\raggedbottom
\relax

\begin{abstract}
\noindent
We update the non-perturbative function of the Collins-Soper-Sterman
(CSS)
resummation formalism which resums the large logarithmic terms
originating
from multiple soft gluon emission in hadron collisions.
Two functional forms in impact parameter ($b$) space are considered,
one with a pure Gaussian form
with two parameters and another with an additional linear term. The
results for the two parameter fit are found to be
$g_1=0.24^{+0.08}_{-0.07}$~GeV$^2$, $g_2=0.34^{+0.07}_{-0.08}$~GeV$^2$.
The results for the three parameter fit are
$g_1=0.15^{+0.04}_{-0.03}$~GeV$^2$, $g_2=0.48^{+0.04}_{-0.05}$~GeV$^2$,
and $g_3=-0.58^{+0.26}_{-0.20}$~GeV$^{-1}$. We also discuss the
potential
of the full Tevatron Run~1
$Z$ boson data for further testing of the universality of the
non-perturbative function.
\\[0.4cm]
\\[0.2cm]
\end{abstract}
\end{center}

\newpage
\setcounter{page}{1}
\pagenumbering{arabic}
\pagestyle{plain}

\setcounter{footnote}{0}
\renewcommand{\thefootnote}{\arabic{footnote}}

\def\doublespaced{\baselineskip=\normalbaselineskip\multiply\baselineskip
by 150\divide\baselineskip by 100}
\doublespaced


\section{Introduction}

It is a prediction of the theory of Quantum Chromodynamics
(QCD) that at hadron colliders the
production of Drell-Yan pairs or weak gauge bosons ($W^\pm$ and $Z$)
will generally be  accompanied by gluon radiation.
Therefore, in order to test QCD theory or the electroweak properties
of vector bosons, it is necessary to include the
effects of multiple gluon emission.
At the Fermilab Tevatron (a $p \bar p$ collider), we expect
about $2\times 10^{6}$ $W^\pm$ and $6 \times 10^{5}$ $Z$ bosons
produced at $\sqrt{S}=1.8$\,TeV, per $100\,{\rm pb}^{-1}$ of
luminosity. This large sample of data is useful
(i) for QCD studies (with either single or multiple scales),
(ii) as a tool for precision measurements of the $W$ boson mass and
width,  and
(iii) as a probe for new physics (e.g., $Z'$).
Achievement of  these  physics goals requires accurate
predictions
for  the distributions of the rapidity and the transverse momentum
of $W^\pm/Z$ bosons and of their decay products.

Consider the production process $h_1 h_2 \rightarrow V X$.
Denote $Q_T$ and $Q$ to be the transverse momentum and the invariant
mass of the vector boson $V$, respectively.
When $Q_T \sim Q$, there is only one hard scale
and a fixed-order perturbation calculation is reliable.
When $Q_T \ll Q$, there are two hard scales and
the convergence of the conventional
perturbative expansion is impaired.
Hence, it is necessary to apply the technique of QCD resummation to
combine the {\it singular} terms in each order of perturbative 
calculation, which yields:
\begin{eqnarray}
\frac{d\sigma}{dQ_T^2} \sim \frac{1}{Q_T^2}
\left\{ \right.
      \alpha_S (L + 1) &+ \alpha_S^2 (L^3 + L^2)
&+ \alpha_S^3 (L^5 + L^4) +
      \alpha_S^4 (L^7 + L^6) + ... \\ \nonumber
                         &+ \alpha_S^2 (~L + ~~1)
&+ \alpha_S^3 (L^3 + L^2) +
      \alpha_S^4 (L^5 + L^4) + ... \\ \nonumber
                   &    &+ \alpha_S^3 (~L + ~~1)
+ \alpha_S^4 (L^3 + L^2) + ...\left. \right\},
\end{eqnarray}
where $\alpha_s$ is the strong coupling constant,
$L$ denotes $\ln(Q^2/Q_T^2)$ and the explicit coefficients
multiplying the logs are suppressed.

Resummation of large logarithms yields a Sudakov form
factor \cite{DDT,Parisi}
and cures divergences as $Q_T \ra 0$.
This resummation was pioneered by Dokshitzer, D'yakonov and Troyan
(DDT)  who performed an analysis in $Q_T$-space which led to a
leading-log resummation formalism \cite{DDT}.
Later, Parisi-Petronzio showed \cite{Parisi} that for
large $Q$ the $Q_T \ra 0$ region can be calculated perturbatively by
imposing the condition of transverse momentum conservation,
\begin{equation}
\delta^{(2)} \left( \sum_{i=1}^{n} {\vec k_{T_i}} - {\vec q_T} \right)
=\int {d^2b \over 4 \pi^2} \;\TT{e}^{i{\vec{Q}_T}\cdot{\vec b}}
\prod_{i=1}^{n} \;\TT{e}^{i{\vec k_{T_i}}\cdot{\vec b}},
\end{equation}
in $b$-space ($b$ is the impact parameter, which is
the Fourier conjugate of $Q_T$).
Their improved formalism also sums some subleading-logs.
They showed that as $Q \ra \infty$,
events at $Q_T \sim 0$ may be obtained asymptotically  by the
emission of  at least two gluons whose transverse momenta are
not small and add to zero.
The intercept at $Q_T=0$ is predicted to be \cite{Parisi}
\begin{equation}
\left.\frac{d\sigma }{dQ_T^2}\right| _{Q_T \ra 0} \sim
\sigma_0 \left( {\Lambda^2 \over Q^2} \right)^{\eta_0},
\end{equation}
\noindent
where $\eta_0=A\ln\left[1+{1\over A}\right]$
with $A=12 C_F/(33-2 n_f)$, and
$\eta_0\simeq 0.6$ for $n_f=4$ and $C_F=4/3$.
Collins and Soper extended \cite{cs} this work in $b$-space
and applied the properties of the renormalization group invariance
to create a formalism that resums all the large log terms
to all orders in $\alpha_s$.

Although various formalisms for resumming large
$\ln(Q^2/Q^2_T)$ terms have been proposed in the literature
\cite{EMP,qtres},
we will concentrate in this paper on the formalism given by
Collins, Soper and Sterman (CSS) \cite{CSS}, which
has been applied to studies of the production of single
\cite{AK,oneW,wres,ERV} and double \cite{twoW}
weak gauge bosons as well as Higgs bosons \cite{Higgs}
at hadron colliders.

\section{Collins-Soper-Sterman Resummation Formalism}

In the CSS resummation formalism,
the cross section is written in the form
\begin{equation}
\label{WY}
{\frac{d\sigma (h_1h_2\rightarrow VX)}{dQ^2\,dQ_T^2dy\,\,}}=
{\frac 1{(2\pi)^2}}\int d^2b\,e^{i{\vec{Q}_T}\cdot {\vec{b}}}
{\widetilde{W}(b,Q,x_1,x_2)}+~Y(Q_T,Q,x_1,x_2),
\end{equation}
where $y$ is the rapidity of the vector boson $V$, and
the parton momentum fractions are defined as $x_1=e^yQ/\sqrt{S}$ and
$x_2=e^{-y}Q/\sqrt{S}$ with $\sqrt{S}$ as the center-of-mass (CM) 
energy of the hadrons $h_1$ and $h_2$.
In Eq.~(\ref{WY}), $Y$ is
the regular piece which can be obtained by subtracting the
singular terms from the exact fixed-order result. The quantity
${\widetilde{W}}$ satisfies a
renormalization group equation with the
 solution of the form
\begin{equation}
{\widetilde{W}(Q,b,x_1,x_2)}=e^{-{\cal S}(Q,b{,C_1,C_2})}
{\widetilde{W}\left(\frac{C_1}{C_2b},b,x_1,x_2\right)}.
\end{equation}
Here the Sudakov exponent is defined as
\begin{equation}
{\cal S}(Q,b{,C_1,C_2})=\int_{C_1^2/b^2}^{C_2^2 Q^2}
\frac{d\overline{\mu }^2}{
\overline{\mu }^2}\left[ {A}\left( {\alpha}_s(\overline{\mu
}),C_1\right)
\ln
\left( \frac{C_2^2 Q^2}{\overline{\mu }^2}\right)
{+B}\left( {\alpha}_s(\overline{\mu }),C_1,C_2\right) \right],
\end{equation}
and the $x_1$ and $x_2$ dependence of $\widetilde{W}$ factorizes as
\begin{equation}
{\widetilde{W}\left(\frac{C_1}{C_2b},b,x_1,x_2\right)}=
\sum_je_j^2\;{\cal C}_{jh_1}\left(\frac{C_1}{C_2b},b,x_1\right)\;
{\cal C}_{jh_2}\left(\frac{C_1}{C_2b},b,x_2\right).
\end{equation}
Here, ${\cal C}_{jh}$ is a convolution of the parton
distribution function ($f_{a/h}$) with calculable Wilson coefficient
functions ($C_{ja}$), which  are defined through
\begin{equation}
{\cal C}_{jh}(Q,b,x)=\sum_{a}\int_x^1{\frac{d\xi }\xi }\;
C_{ja}\left(\frac{x}{\xi},b,\mu={C_3 \over b},Q\right)\;
f_{a/h}\left(\xi ,\mu={C_3 \over b} \right).
\end{equation}
The sum on the index $a$ is over incoming partons, $j$ denotes the quark
flavors with (electroweak) charge $e_j$, and the factorization scale $\mu$
is fixed to be $C_3/b$.
A few comments about this formalism:
\begin{itemize}
\item
The $A$, $B$ and $C$ functions can be calculated order-by-order
in $\alpha_s$.
\item
A special choice can be made for the renormalization constants $C_i$
 so that the contributions obtained from the expansion in $\alpha_s$
of the CSS resummed calculation agree with those from the fixed-order
calculation. This is the canonical choice. It has
$C_1=C_3=2e^{-\gamma _E}\equiv b_0$ and $C_2=C_1/b_0=1$,
where $\gamma _E$ is Euler's constant.
\item
$b$ is integrated from 0 to $\infty$.
For  $b\gg 1/\Lambda_{QCD}$, the perturbative calculation is
no longer reliable.
In order to account for  non-perturbative contributions from the large
$b$
region this formalism includes an additional
multiplicative factor which contains measurable parameters.
\end{itemize}
We refer the readers to Ref. \cite{wres} for a more detailed discussion
on how to apply the CSS resummation formalism
to the phenomenology of hadron collider physics.

\subsection{ The Non-perturbative Function}

As noted in the previous section, it is necessary to include an
additional factor, usually referred to as the
``non-perturbative function'', in the CSS resummation formalism in
order to  incorporate some long distance physics not accounted for by
the
perturbative derivation.
Collins and Soper postulated \cite{CSS}
\begin {equation}
\widetilde{W}_{j{\bar{k}}}(b)=
\widetilde{W}_{j{\bar{k}}}(b_{*})\widetilde{W}
_{j{\bar{k}}}^{NP}(b)\,,
\end{equation}
with
\begin{equation}
b_{*}={\frac b{\sqrt{1+(b/b_{max})^2}}}\,,
\end{equation}
so that $b$ never exceeds $b_{max}$ and
$\widetilde{W}_{j{\bar{k}}}(b_{*})$ can be reliably calculated
in perturbation theory.
(In numerical calculations, $b_{max}$ is typically set to be
of the order of $1~{\rm GeV}^{-1}$.)
Based upon a renormalization group analysis, they found that
the non-perturbative function can be generally written as
\begin{equation}
\widetilde{W}_{j\bar{k}}^{NP}(b,Q,Q_0,x_1,x_2)=\exp \left[ -F_1(b)\ln
\left( \frac{Q^2}{Q_0^2}\right) -F_{j/{h_1}}(x_1,b)-F_{{\bar{k}}/{h_2}
}(x_2,b)\right]\,,
\end{equation}
where $F_1$, $F_{j/{h_1}}$ and $F_{{\bar{k}}/{h_2}}$ must be
extracted from data with the constraint that
\bea
\widetilde{W}_{j\bar{k}}^{NP}(b=0)=1.
\ena
Furthermore,
$F_1$ only depends on $Q$, while
$F_{j/{h_1}}$ and $F_{{\bar{k}}/{h_2}}$ in general depend on
$x_1$ or $x_2$, and their values can depend on the flavor of the
initial state partons ($j$ and $\bar k$ in this case).
Later, in Ref. \cite{sterman}, it was shown that the
$F_1(b)\ln \left( \frac{Q^2}{Q_0^2}\right)$
dependence is also suggested by
infrared renormalon contributions to the $Q_T$ distribution.

\subsection{Testing the Universality of
$\widetilde{W}_{j\bar{k}}^{NP}$}

The CSS resummation formalism suggests that the non-perturbative
function
should be universal. Its role is analogous to that of the parton
distribution  function (PDF) in any fixed order perturbative
calculation,
as its value must be determined from data.
The first attempt to determine such a universal non-perturbative
function
was made by Davies, Webber and Stirling (DWS)~\cite{DWS}
in 1985 who fit data available at that time to the resummed piece 
(the $\widetilde{W}_{j\bar{k}}$ term)
Duke and Owens parton distribution functions
\cite{dopdf}. Subsequently, the DWS results were combined with a NLO
calculation \cite{ArnoldReno} by Arnold and Kauffman
\cite{AK} in 1991 to provide the first complete CSS prediction
relevant to hadron collider Drell-Yan data. In 1994, Ladinsky and Yuan
(LY)~\cite{LY} observed that the prediction  of the DWS set of
$\widetilde{W}_{j\bar{k}}^{NP}$   deviates from R209 data
($p+p \ra \mu^+ \mu^- + X$ at $\sqrt{S}=62$\,GeV) using the
CTEQ2M PDF \cite{cteq2}.
In order to incorporate possible $\ln(\tau)$ dependence,
LY postulated a model for the non-perturbative term, which was different
from that of DWS, as
\begin{equation}
\widetilde{W}_{j\bar{k}}^{NP}(b,Q,Q_0,x_1,x_2)={\rm exp}\left[
-g_1b^2-g_2b^2\ln \left( {\frac Q{2Q_0}}\right) -g_1g_3b\ln
{(100x_1x_2)}
\right] ,
\end{equation}
where $x_1x_2=\tau$.
A ``two-stage fit'' of the R209, CDF-$Z$ ($4\,{\rm pb}^{-1}$ data)
and E288 ($p+Cu$) data gave \cite{LY}
$$
g_1=0.11_{-0.03}^{+0.04}~{\rm GeV}^2\; , \;
g_2=0.58_{-0.2}^{+0.1}~{\rm GeV}^2 \; , \;
g_3=-1.5_{-0.1}^{+0.1}~{\rm GeV}^{-1},
$$
for $Q_0=1.6~{\rm GeV}$ and
$b_{max}=0.5~{\rm GeV}^{-1}$.
\footnote{A {\tt {\small FORTRAN}} coding error in calculating the
parton densities of the neutron
led to an incorrect value for $g_3$.}
The purpose of the project described here is to update these
non-perturbative
parameters using modern, high-statistics samples of Drell-Yan data and
to
incorporate a  fitting technique which will track the full error matrix
for all  fitted parameters. Our results are given in
the following sections.

\section{Fitting Procedure}

\subsection{Choice of the Parametrization Form}

At the present time, the non-perturbative functions in the
CSS resummation formalism cannot be derived from QCD theory,
so a variety of functional forms should be studied.
The only necessary condition is that
$\widetilde{W}_{j\bar{k}}^{NP}(b=0)=1$. For simplicity, we consider only
two typical functional forms for
$\widetilde{W}_{j\bar{k}}^{NP}(b,Q,Q_0,x_1,x_2)$ in $b$ space:
(i) 2-parameter pure Gaussian form [DWS form]:
\begin{equation}
{\rm exp}\left[-g_1-g_2\ln \left( {\frac Q{2Q_0}} \right)
\right] b^2 \, ,
\end{equation}
with $Q_0=1.6$\,GeV; and
(ii) a 3-parameter form [LY form]:
\begin{equation}
{\rm exp}\left\{ \left[
-g_1-g_2\ln \left( {\frac Q{2Q_0}} \right) \right] b^2
-\left[ g_1g_3\ln\!{(100x_1x_2)} \right] b
\right\} \, ,
\label{LY_form}
\end{equation}
with a logarithamic $x$-dependent third term which is linear in $b$.
This is equivalent to $\ln({\tau}/{\tau_0})$ for
$\tau_0=0.1$.

Both forms assume no flavor dependence for simplicity. In addition to
fitting for the non-perturbative parameters, $g_1,\, g_2,\, \mbox
{and}\,
g_3$, the overall normalizations were allowed to float for some
fits. One can also study another pure Gaussian form with similar $x$
dependence such as
\bea
{\rm exp}\left[
-g_1-g_2\ln \left( {\frac Q{2Q_0}} \right)
-g_1g_3\ln {(100x_1x_2)} \right] b^2
\, .
\ena
However, we find that current data are not yet precise enough to
clearly separate the $g_2$ and $g_3$ parameters within this
functional form and so it is not considered here.
We also tested a few additional functions which did not incorporate
additional
parameters, but did not find any clear advantage to them when fitting the
current Drell-Yan data. However, as to be shown later, the Run~1 $W$/$Z$
data  at the Tevatron are expected to determine the $g_2$ coefficient
with good accuracy, and these data can be combined with the low
energy Drell-Yan data to further test various scenarios for
$x$ dependence and ultimately, universality.

\subsection{Choice of the Data Sets}

In order to determine the non-perturbative functions discussed above,
we need to choose  experimental data sets for which the contribution
to the non-perturbative piece dominates the
transverse momentum distributions.
This suggests using low energy fixed target or collider
data in which the transverse momentum ($Q_T$)
of the Drell-Yan pair is much less than its invariant mass ($Q$).
Because the CSS resummation formalism better describes
data in which the Drell-Yan pairs are produced in the central rapidity
region (as defined in the center-of-mass frame of the
initial state hadrons)
we shall concentrate on data with those properties.
Based upon the above criteria we chose to consider data shown in
Table~\ref{data_table}.
\begin{table}[t]
\begin{center}
\begin{tabular}{ c | c | c | c | c | c}
 Experiment & reference & Reaction & $\sqrt{S}$~GeV & $\langle \tau
\rangle$ & $\delta N$ \\
\hline
R209 & \cite{R209} & $p+p \ra \mu^+ \mu^- +X$ & 62 & $\sim 0.1$ 
& 10\% \\
\hline
E605 & \cite{E605} & $p+Cu \ra \mu^+ \mu^- +X$ & 38.8 & $\sim 0.2$ 
& 15\% \\
\hline
CDF-$Z$ & \cite{CDFZ} & $p + \bar p \ra Z +X$ & 1800 & $\sim 0.05$ 
& --  \\
\hline
E288 & \cite{E288} & $p+Cu \ra \mu^+ \mu^- +X$ & 27.4 & $\sim 0.2$ 
& 25\% \\
\end{tabular}
\caption[]{Drell-Yan data used in this analysis. Here, $\delta N$ is the
published normalization uncertainty for each experiment. The CDF data
were
from Tevatron collider Run 0 of 4 pb$^{-1}$.}
\label{data_table}
\end{center}
\end{table}
We have also examined the E772 data \cite{E772},
from the process
$p+ H^2 \ra \mu^+ \mu^- +X$ at $\sqrt{S}=56.6$\,GeV,
and found that it was not compatible
in our fits with the above data,
and it is not included in this study.\footnote{ Using the fitted
$g$ values to be given below, the theory prediction for the E772
experiment is typically a factor of 2 smaller than the data. Similarly,
CTEQ fitting of PDF parameters are not well fit with these data \cite{wkt}.}
Except where noted, all of
the fits to $g_{1,2,3}$ were done using the CTEQ3M PDF \cite{cteq3}
fits.
\begin{figure}
\centerline{
\psfig{figure=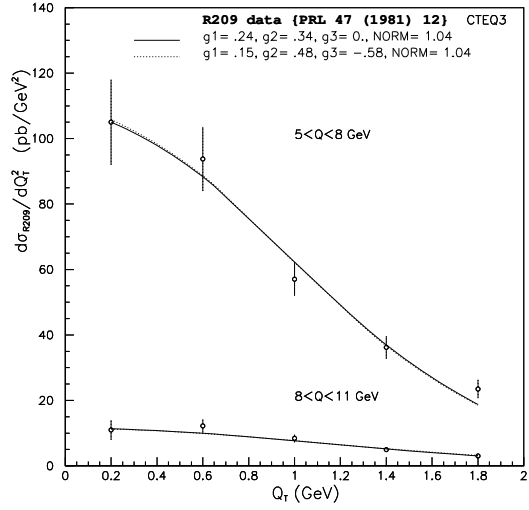,height=3.in}
}
\caption{R209 data, from $p+p \ra \mu^+ \mu^- +X$ at
$\protect\sqrt{S}=62$\,GeV,
with an overall systematic normalization error of $10\%$. The curves are
the results of Fit $A_2$ and $A_3$ and are multiplied by the
value of NORM, as shown in  the figure and described in the text.}
\label{fig:r209}
\end{figure}
\begin{figure}
\centerline{
\psfig{figure=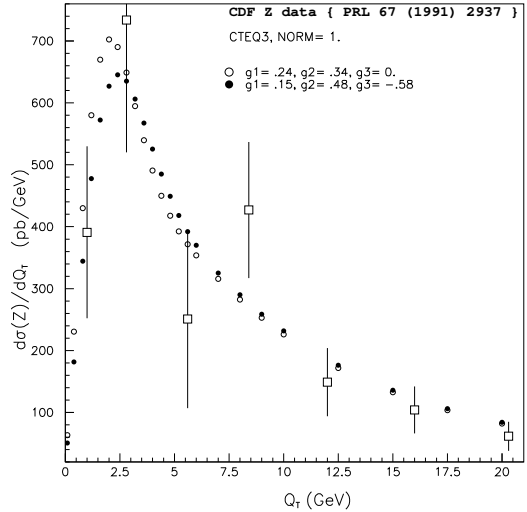,height=3.in}
}
\caption{Comparison of $4\,{\rm pb}^{-1}$ CDF-$Z$ data at the Tevatron
with two different theory model predictions. The dots correspond to the
results of Fit $A_2$ and $A_3$.}
\label{fig:cdfZ}
\end{figure}
\begin{figure}
\centerline{
\psfig{figure=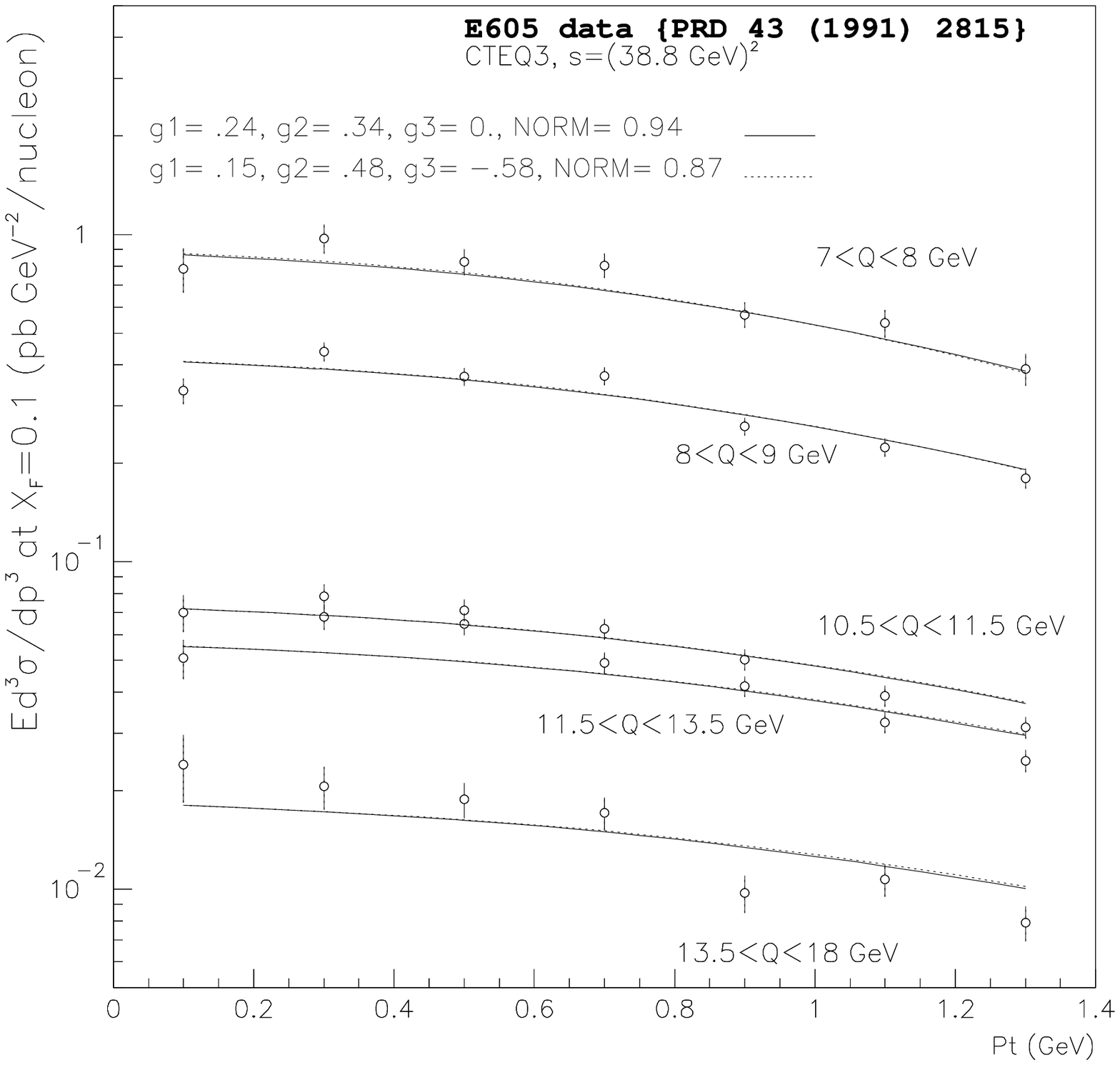,height=3.in}
}
\caption{E605 data, from $p+Cu \ra \mu^+ \mu^- +X$ at
$\protect\sqrt{S}=38.8$\,GeV,
with an overall systematic normalization error of $15\%$.
 The curves are
the results of Fit $A_2$ and $A_3$ and are multiplied by the
value of NORM, as shown in  the figure and described in the text.}
\label{fig:e605}
\end{figure}

\subsection{Primary Fits}

As to be shown later,
the E288 data have the
smallest errors, and
would be expected to dominate the result of a global fit.
That is indeed the case. When including the
E288 data in a global fit, we found that the  resulting fit required the
NORM\footnote{Here the quantity NORM is the fitted normalization factor
which is applied to the prediction curves in all that follow: the data
are uncorrected.} to be too large (as compared to the experimental
systematic error) for either the E288 or the E605 data.
Furthermore, the shape of the R209 data cannot be well described by the
theory prediction based on such a fit. 

\subsubsection{Fits $A_{2,3}$}

We therefore employed a different strategy for the global fit
based on the statistical quality of the data. We included the first two
mass bins  ($7<Q<8$\,GeV and
$8<Q<9$\,GeV) of the E605 data and all of the R209 and the CDF-$Z$ boson
data, in an initial global fit, referred to here as Fit $A_{2,3}$.
In total, 31 data points were considered.
 We allowed the normalization
of the R209 and E605 data to float within their overall systematic
normalization errors, while fixing the normalization
of the CDF-$Z$ data to unity.
(The point-to-point systematic error of 10\% for the E605 data has
also been included in the error bars of the
data points shown in Fig.~\ref{fig:e605}.)
In addition to the normalization factor for each
experiment, the fitted parameters of our global fit include
the coefficients $g_{1,2}$ and $g_{1,2,3}$ for the 2-parameter
and 3-parameter fits, respectively.
We found that, for $Q_0=1.6$\,GeV and $b_{max}=1/(2\,{\rm GeV})$,
both the 2-parameter and the 3-parameter forms give good fits, with
$\chi^2$ per degree of freedom about equal to 1.4.\footnote{
We scan the values of $g_1$ and $g_2$ between 0 and 1, and $g_3$ between
$-2$ and 3.}
The best fitted central values for Fit $A_2$ are:
$g_{1}  =  0.24~{\rm GeV}^2\; , \; g_{2} = 034~{\rm GeV}^2$.
While the central values for Fit $A_{3}$ are
$g_{1}  =  0.15~{\rm GeV}^2\; , \; g_{2} = 0.48~{\rm GeV}^2\; , \;
g_{3} = -0.58~{\rm GeV}^{-1}$. The fitted values for NORM are 1.04 for
both Fits $A_2$ and $A_3$.

\subsubsection{Uncertainties in the Fits}
\begin{figure}
\centerline{
\psfig{figure=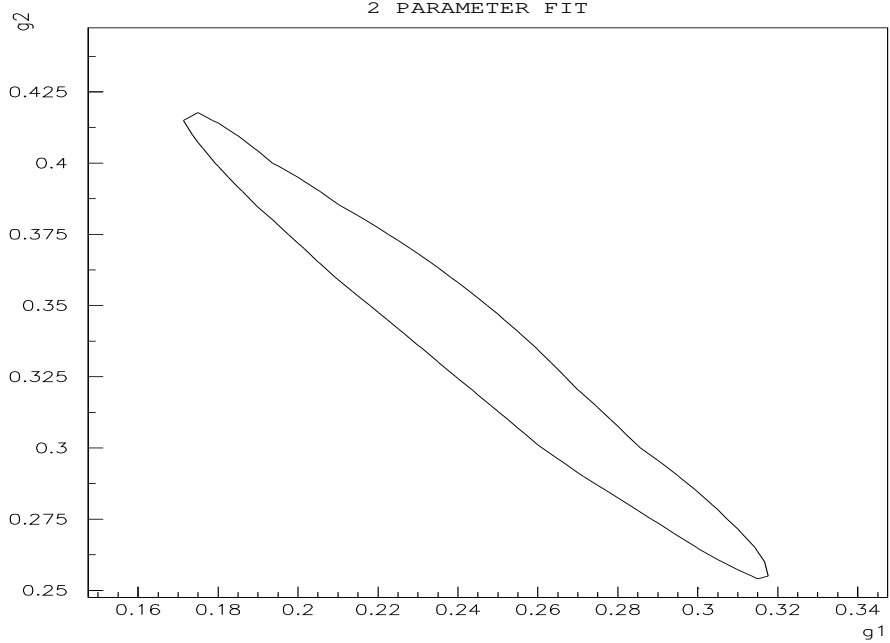,width=12cm,height=3.5in}
}
\caption{ The error ellipse on the $g_1$ and $g_2$ plane
from which 
the errors of the 2-parameter fit $A_2$ were interpreted.
}
\label{fig:cont2}
\end{figure}
\begin{figure}
\begin{center}
\vspace*{-0.8cm}
\begin{tabular}{cc} 
\psfig{figure=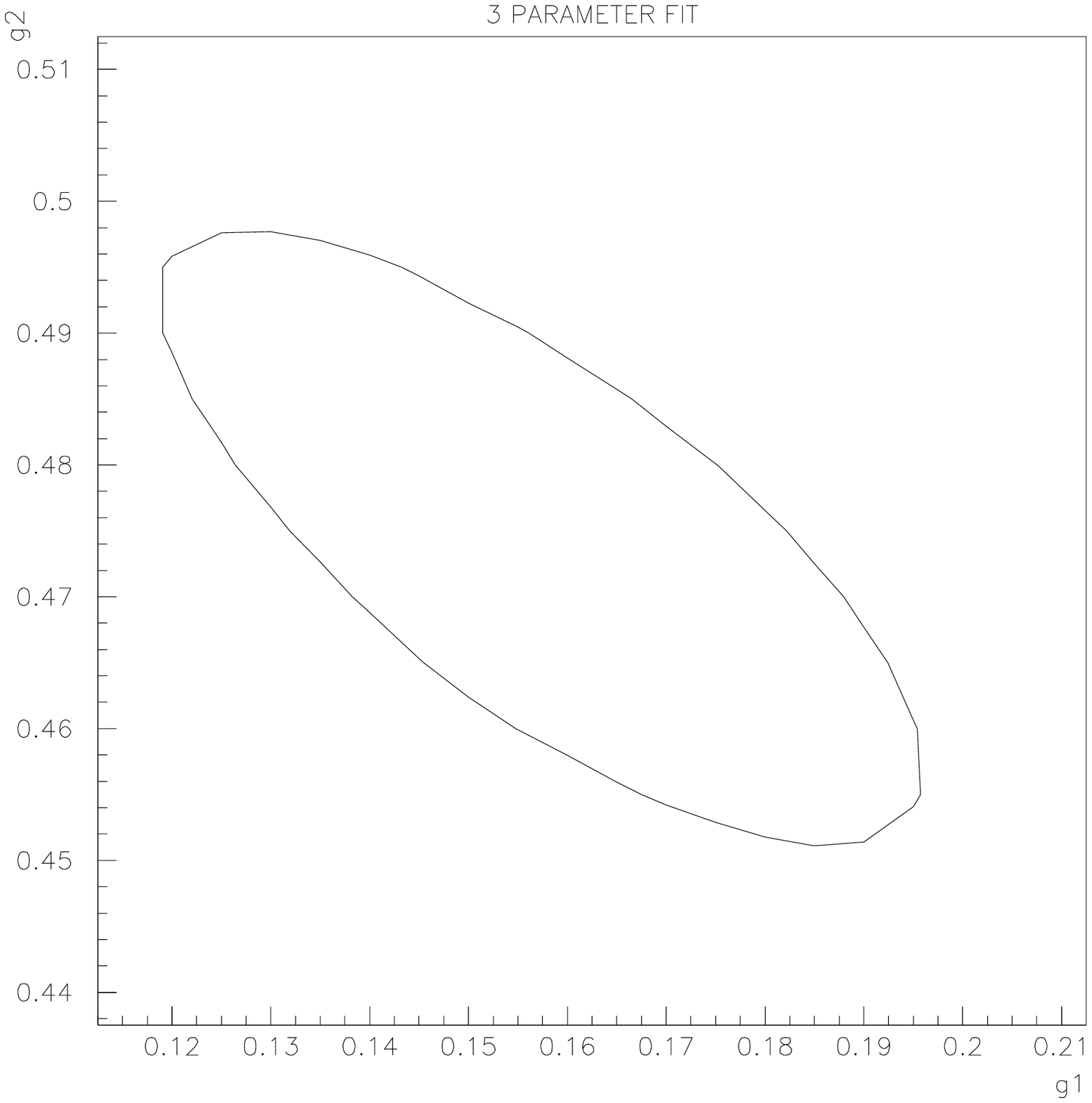,width=12cm,height=3.0in} 
\\[-1.5in]\hskip 4.9in (a)\\[0.9in]
\psfig{figure=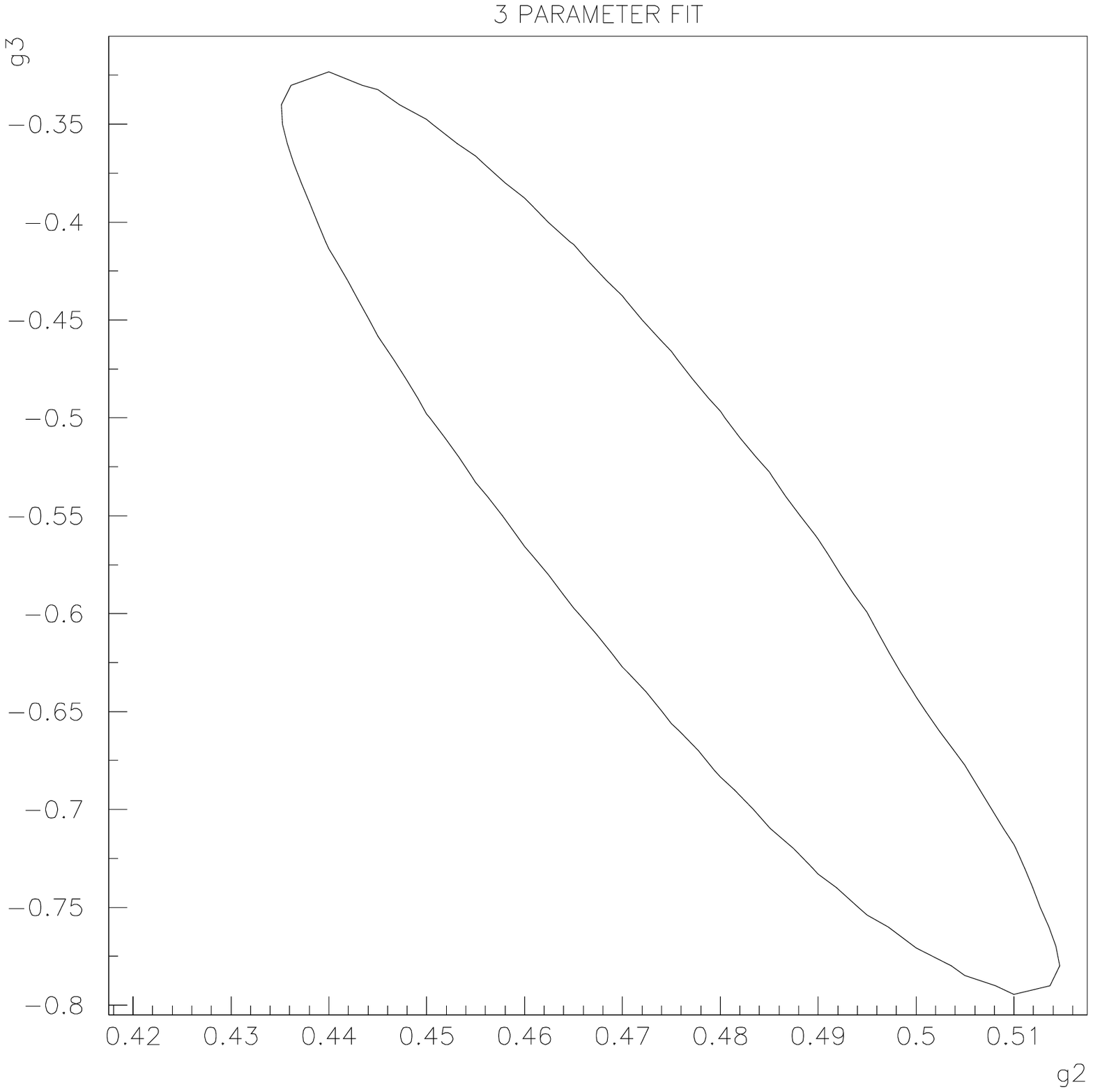,width=12cm,height=3.0in} 
\\[-1.5in]\hskip 4.9in (b)\\[0.9in]
\psfig{figure=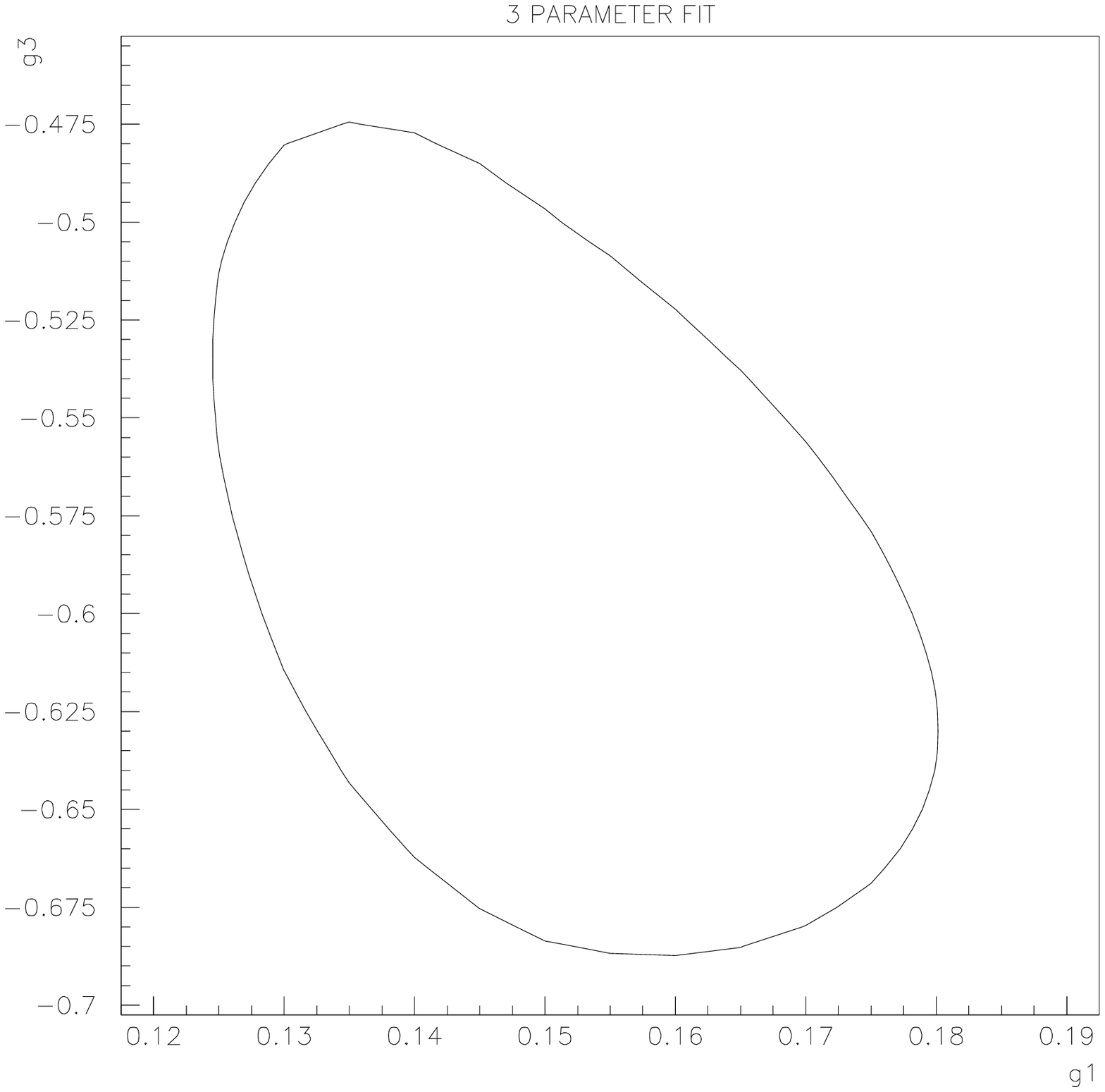,width=12cm,height=3.0in} 
\\[-1.5in]\hskip 4.9in (c)\\[0.9in]
\end{tabular}
\end{center}
\caption{ The error ellipse projections from which  
the errors of the 3-parameter fit $A_3$ were interpreted. (a) $g_1$ and
$g_2$ plane, (b) $g_2$ and $g_3$ plane, and (c) $g_1$ and $g_3$ plane.
}
\label{fig:cont3}
\end{figure}

We have also studied the uncertainties of the fitted
$g$ parameters.
For the 2-parameter fit, the $1\sigma$ error in the $\chi^2$ plot
(with an approximately elliptical contour) gives
$ -0.07 < \delta g_1 < 0.08~{\rm GeV}^2 $
and
$ -0.08 < \delta g_2 < 0.07~{\rm GeV}^2 $.
For the 3-parameter fit, the situation is more complicated,
as the fitted values of $g$'s are highly correlated.
In order to estimate the uncertainties of the fitted $g$ values, we
fix the value (at its best fit value) of $g$'s, one at a time,
and examine the uncertainties of the other two,
in a way similar to studying the 2-parameter fit result.
We found that
\begin{eqnarray*}
\mbox{$g_1$ fixed:} -0.05 & < & \delta g_2 < 0.04~{\rm GeV}^2 \; , \\
                    -0.02 & < & \delta g_3 < 0.26~{\rm GeV}^{-1} \; ;\\
\mbox{$g_2$ fixed:} -0.02 & < & \delta g_1 < 0.03~{\rm GeV}^2 \; , \\
                    -0.08 & < & \delta g_3 < 0.07~{\rm GeV}^{-1} \; ; \\
\mbox{$g_3$ fixed:} -0.02 & < & \delta g_1 < 0.04~{\rm GeV}^2 \; ,\\
                    -0.03 & < & \delta g_2 < 0.02~{\rm GeV}^2 \; ;
\end{eqnarray*}
constitute a conservative set of uncertainty ranges.
With the understanding of the complexity discussed above, we
characterize the uncertainties of the fitted $g$'s in the 3-parameter
form conservatively by their maximal deviations, so
that the best fitted $g$ values are:
\begin{eqnarray*}
{\rm Fit }\;A_{2}:\\
   g_{1} & = & 0.24^{+0.08}_{-0.07}~{\rm GeV}^2\; , \; g_{2} =
034^{+0.07}_{-0.08}~{\rm GeV}^2 \; ; \\
{\rm Fit }\;A_{3}:\\
   g_{1} & = & 0.15^{+0.04}_{-0.03}~{\rm GeV}^2\; , \; g_{2} =
0.48^{+0.04}_{-0.05}~{\rm GeV}^2\; , \; g_{3} =
-0.58^{+0.26}_{-0.20}~{\rm GeV}^{-1} \; ;
\end{eqnarray*}
for the 2 and 3 parameter fits, respectively.
In Figs.~\ref{fig:cont2} and \ref{fig:cont3},
we show the error ellipse projections from which we interpreted the 
errors of the above fits.
In summary, fits $A_2$ and
$A_3$ constitute the main results of this paper.

\subsubsection{Cross Checks}
\begin{figure}
\centerline{
\psfig{figure=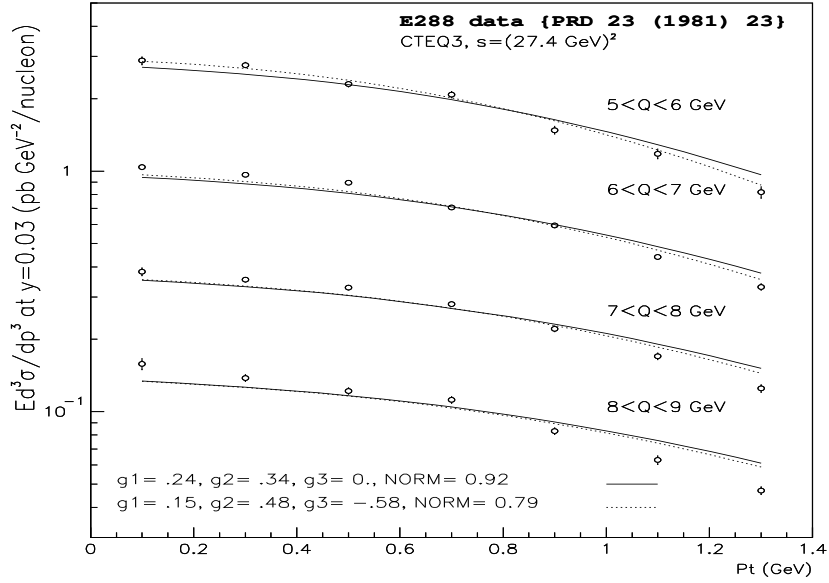,width=12cm,height=3.5in}
}
\caption{E288 data, from $p+Cu \ra \mu^+ \mu^- +X$ at
$\protect\sqrt{S}=27.4$\,GeV,
with an overall systematic normalization error of $25\%$.
 The curves are
the results of Fit $N_1$ and $N_2$ and are multiplied by the
value of NORM, as shown in  the figure and described in the text.}
\label{fig:e288}
\end{figure}

Given these values of $g$'s and the fitted normalization factors for
the E605 data, we can calculate the two different predictions
for the other three high mass bins not used in Fits $A_{2,3}$,
and the results are also shown in Fig.~\ref{fig:e605}.
In order to compare with the E288 data, we created Fits $N_{2,3}$
in which we fix the $g$'s to those obtained from Fit $A_3$ and fit for
NORM from the E288 data alone. Fig.~\ref{fig:e288} shows the resulting
fits are acceptable, with values of NORM close to the quoted 25\%,
namely, NORM$=0.92$ and $0.79$ for Fits $N_{2,3}$,
respectively. It is encouraging that the quality of the fit for the E288
results is very similar to that for the E605 data and that the
normalizations are now acceptably within the range quoted by the
experiment. Hence, we conclude that the fitted values of
$g$'s reasonably describe the wide-ranging, complete set of data, as
discussed above.

We note that
although the CDF-$Z$ data, as shown in Fig.~\ref{fig:cdfZ},
contain only 7 data points with large statistical uncertainties, they
prove to be very useful in determining the value of $g_2$.
To test this observation, we performed an additional fit, $L_2$.
Following the method suggested in \cite{LY}, we set $g_3$ to be zero and
fit the $g_1$ and $g_2$ parameters using the R209 and CDF-$Z$ data alone.
(Note that for the R209 data, the typical value of $\tau$
is of the order 0.1, which motivates the choice of $\tau_0 = 0.1$ in
the LY form. Effectively, the $g_3$ contribution to the R209 data can be
ignored.) We found that the best fit\footnote{Also, $g_1=0.18~{\rm GeV}^2$.}
gives $g_2=0.47~{\rm GeV}^2$,
 which is in good agreement with the result of the global Fit $A_3$ 
discussed above.
Hence, we conclude that the CDF-$Z$ data already play an important
role in constraining the $g_2$ parameter, which can be further improved
with a large $Z$ data samples from Run~1 of the Tevatron collider
experiments. We shall defer its discussion to the next section.

\section{Run~1 $W$ and $Z$ Boson Data at the Tevatron}

The Run~1 $W$ and $Z$ boson data at the Tevatron can be useful as a
test of {universality} and the $x$ dependence of
the non-perturbative function
$\widetilde{W}_{j\bar{k}}^{NP}(b,Q,Q_0,x_1,x_2)$.
This is clearly demonstrated in Fig.~\ref{fig:cdfZ},
where we give the predictions for the two different
global fits (2-parameter and 3-parameter fits)
obtained in the previous section using the CTEQ3M PDF parameterizations.
(The CTEQ4M PDF \cite{cteq4} gives similar results.)
With the large $Z$ boson data sets anticipated
from Tevatron Run~1 (1a and 1b),
it should be possible to distinguish these two example models.
\begin{figure}
\centerline{
\psfig{figure=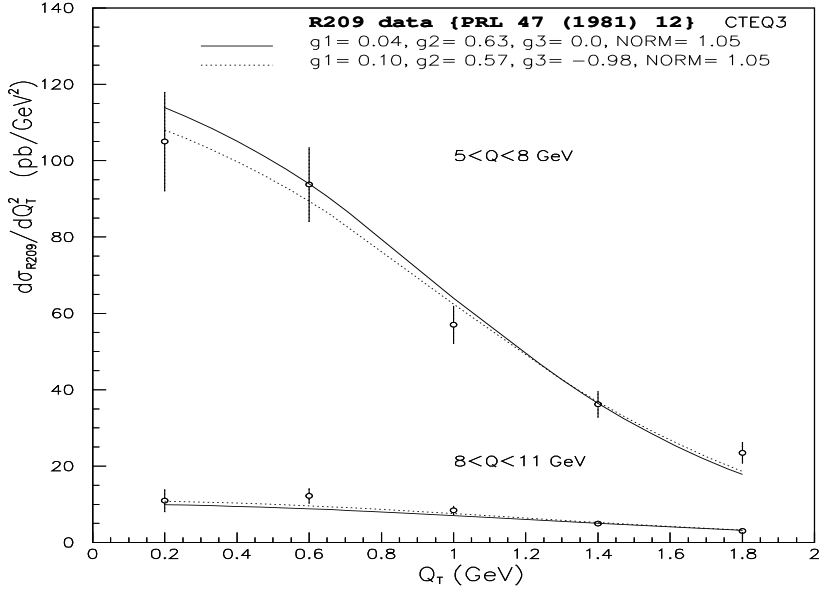,width=12cm,height=3.5in}
}
\caption{ Comparison of R209 data with two different theory
model predictions obtained from the ``toy global fit''.  The curves are
the results of Fit $F_1$ and $F_2$ and are multiplied by the
value of NORM, as shown in  the figure and described in the text.}
\label{fig:r209N}
\end{figure}
\begin{figure}
\centerline{
\psfig{figure=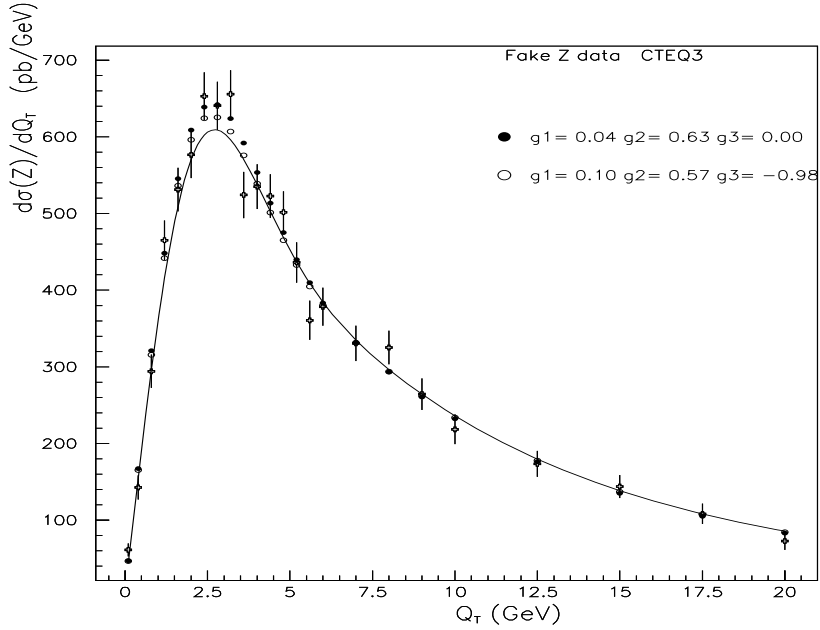,width=12cm,height=3.5in}
}
\caption{ Comparison of ``fake $Z$'' data (solid curve)
with two different theory model predictions (dots) 
obtained from the ``toy global fit'' $F_1$ and $F_2$.}
\label{fig:fakeZ}
\end{figure}
\begin{figure}
\centerline{
\psfig{figure=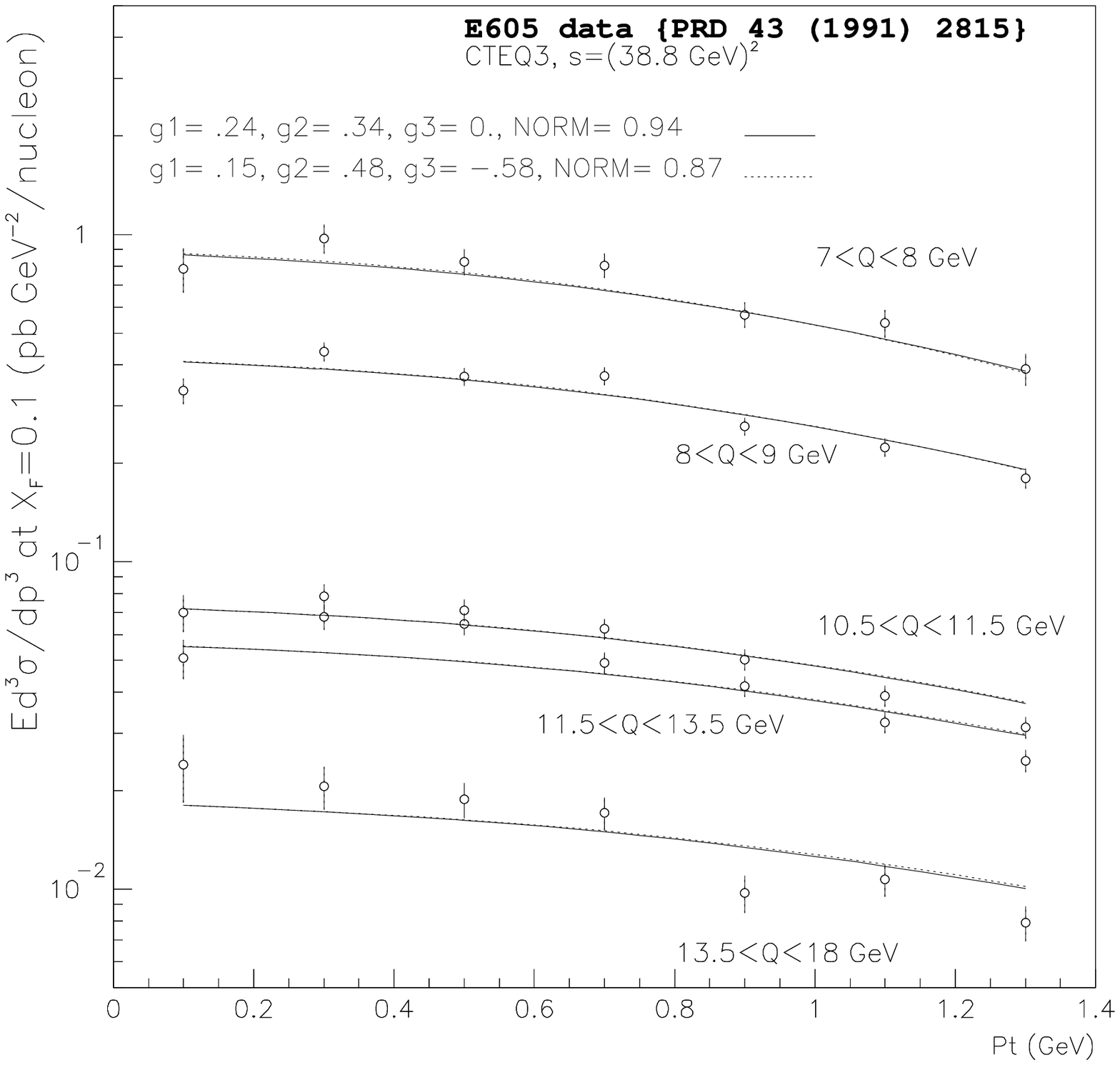,height=3.in}
}
\caption{ Comparison of E605 data with two different theory
model predictions obtained from the ``toy global fit''.  The curves are
the results of Fit $F_1$ and $F_2$ and are multiplied by the
value of NORM, as shown in  the figure and described in the text.}
\label{fig:e605N}
\end{figure}
\begin{figure}
\centerline{
\psfig{figure=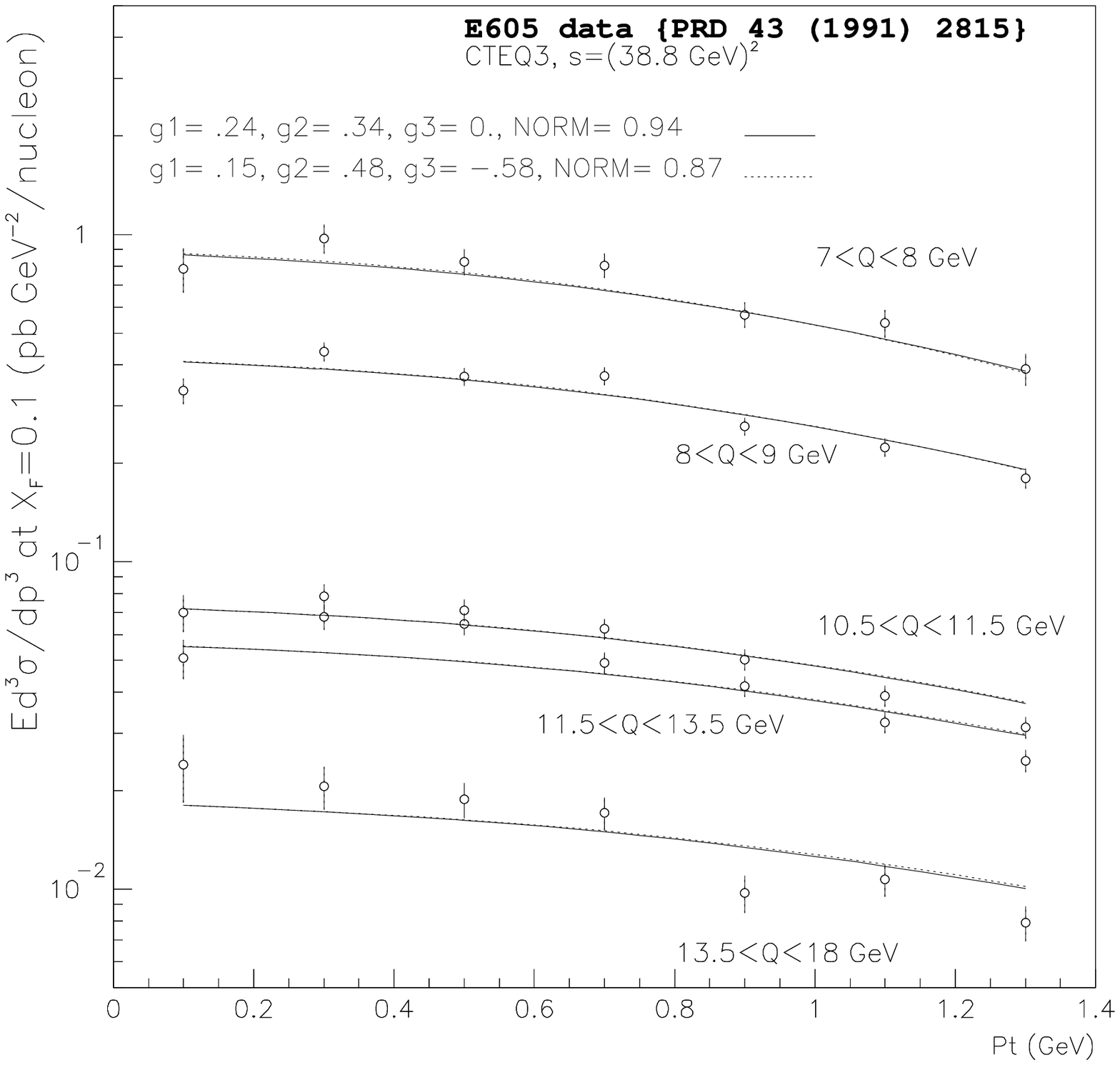,height=3.in}
}
\caption{ Comparison of E288 data with two different theory
model predictions obtained from the ``toy global fit''.  The curves are
the results of Fit $F_1$ and $F_2$ and are multiplied by the
value of NORM, as shown in  the figure and described in the text.}
\label{fig:e288N}
\end{figure}

As shown in Ref.~\cite{wres}, for $Q_T > 10$\,GeV
the non-perturbative function has little effect on the $Q_T$
distribution, although in principle it affects the
whole $Q_T$ range (up to $Q_T \sim Q$).
In order to study the resolving power of the full Tevatron
Run~1 $Z$ boson data in determining the non-perturbative function,
we have performed a ``toy global fit'', Fit $F_3$ as follows.
First, we generate a set of fake Run~1 $Z$ boson data
(assuming 5,500 reconstructed $Z$ bosons in 24 $Q_T$
bins between $Q_T = 0 {\rm and}\; 20$~GeV/$c$) using the original
LY fit results ($g_1=0.11~{\rm GeV}^2$, $g_2=0.58~{\rm GeV}^2$ and
$g_3=-1.5~{\rm GeV}^{-1}$). Then, we combine these fake-$Z$ boson data
with the   R209 and E605 Drell-Yan data as discussed above to perform a
global fit. The 3-parameter form results in
$$
{\rm Fit}\; F_3: \\
g_1 = 0.10^{+0.02}_{-0.02}~{\rm GeV}^2\; , \;
g_2 = 0.57^{+0.01}_{-0.02}~{\rm GeV}^2\; , \;
g_3 =  -0.98^{+0.15}_{-0.17}~{\rm GeV}^{-1} \; ,
$$
with a $\chi^{2}$ per
degree of freedom  of approximately 1.3.\footnote{
This amounts to a shift in the prediction for the mass and the width of
the $W$ boson by about 5\,MeV and 10\,MeV, respectively \cite{new}.}
These fitted values for the $g$'s agree perfectly with those used to
generate the fake-$Z$ data except for the value of
$g_3$, which is smaller by a factor of 2.
It is interesting to note that this result agrees within $2\sigma$
with that of Fit $A_3$,
using only the current low energy data, although
the uncertainties on $g$'s are smaller by a factor of 2.

We have also performed the same fit for the 2-parameter form and
obtained an equally good fit with
$$
{\rm Fit}\; F_2: \\
g_1=0.04^{+0.03}_{-0.03}~{\rm GeV}^2 \; , \;
g_2=0.63^{+0.04}_{-0.03}~{\rm GeV}^2.
$$
While the new $g_2$ value ($0.63~{\rm GeV}^2$)
obtained in Fit $F_2$ is very
different from that of $A_2$ ($0.34~{\rm GeV}^2$) given in the previous
section, it is  actually in good agreement with the $g_2$ value
($0.57~{\rm GeV}^2$) obtained from Fit $A_3$.
This implies that the Run~1 $Z$ boson data,
when combined with the low energy Drell-Yan data,
can be extremely useful in determining the parameter $g_2$.
In Figs.~\ref{fig:r209N}-\ref{fig:e288N}, we
compare the two theory predictions (derived from $F_{2,3}$)
with the R209, fake-$Z$, E605, and E288 data.
As shown, they both agree well with all of the data.

Before closing this section, we would like to comment on
the result of a single parameter study of the fake $Z$ data, Fit $S_1$.
Given the large sample of Run~1 $Z$ data, one can consider
fitting the non-perturbative function with only one, $Q$-independent,
non-perturbative parameter.
With this in mind, we fitted
the fake $Z$ data with the non-perturbative function
\begin{equation}
\widetilde{W}_{j\bar{k}}^{NP}(b,Q,Q_0,x_1,x_2)=
{\rm exp}\left[ - \overline{g_1} b^2 \right] \,,
\end{equation}
and found that 
$$
{\rm Fit}\; S_1: \\
\overline{g_1}=2.1^{+0.09}_{-0.08}~{\rm GeV}^2 \, , 
$$
which gives a good description of the $Q_T$ distribution
of the ``fake-$Z$'' data.
It is obvious that this fitted value
agrees with that of the 2-parameter fit just by considering the
coefficient of $b^2$ in first two terms of Eq.~\ref{LY_form}. For the
results of $F_2$ with the value of the $Z$ boson mass,
$M_Z=91.187$\,GeV$/c^2$, we obtain
$0.04+0.63 \ln(M_Z/2 Q_0)=2.14$, which is essentially the same as the
coeficient of $b^2$ in $S_1$ . One interesting question is whether the
result of this one-parameter fit alone can be used to also describe the
$Q_T$ distribution of the $W^\pm$ boson produced at the Tevatron (at the
same energy). A quantitative estimate can be easily obtained by noting
again that the  difference between
$0.04+0.63 \ln(M_Z/2 Q_0)=2.14$ and
$0.04+0.63 \ln(M_W/2 Q_0)=2.06$ with the $W$-boson mass
$M_W=80.3$\,GeV$/c^2$, is 0.08, which are essentially the same, given
the
uncertainty of $0.09~{\rm GeV}^2$ from $S_1$. We
conclude  that it is indeed a good approximation to use the
one-parameter
fit result from fitting $Z$ boson data in
order to predict the
$Q_T$ distribution of the $W^\pm$ boson using the CSS resummation
formalism. On the other hand,
a single parameter without $Q$ dependence
(i.e. the parameter $\overline{g_1}$ alone)
does not give a reasonable global fit to all of the Drell-Yan data
discussed above. For instance,
for the R209 data, the 2-parameter fit gives
$0.04+0.63 \ln(8/2 Q_0)=0.6$ for the coefficient of Eq.~\ref{LY_form},
which is not consistent with the value of
$\overline{g_1}$ from $S_1$. Hence, we conclude that in order to test
the universality of the non-perturbative function of the CSS
formalism, one must consider
its functional form with $Q$ (and $x$) dependence.

\section{Conclusions}

The effects of QCD gluon resummation are important in many precision
measurements.
In order to make predictions using the CSS resummation formalism for the
$Q_T$ distributions of vector bosons
at hadron colliders,
it is necessary to include contributions from
the phenomenological non-perturbative functions inherent to the
formalism. In this paper, we have extended previous results by making
use of  2-parameter and the 3-parameter fits to modern, low energy
Drell-Yan data.  We found that both parameterizations result in good
fits. In particular our results are
\begin{eqnarray*}
{\rm Fit }\;A_{2}:\\
   g_{1} & = & 0.24^{+0.08}_{-0.07}~{\rm GeV}^2\; , \; g_{2} =
0.34^{+0.07}_{-0.08}~{\rm GeV}^2  \; . \\
{\rm Fit }\;A_{3}:\\
   g_{1} & = & 0.15^{+0.04}_{-0.03}~{\rm GeV}^2\; , \; g_{2} =
0.48^{+0.04}_{-0.05}~{\rm GeV}^2\; , \; g_{3} =
-0.58^{+0.26}_{-0.20}~{\rm GeV}^{-1} \; .
\end{eqnarray*}
Each functional form predicts measurably different $Q_T$
distributions for $Z$ bosons  produced at the Tevatron.
We showed that the full Tevatron Run~1 $Z$ boson data can
potentially distinguish these two different models.

In particular, using the results from a toy global fit, we concluded
that
the large sample of the Run~1 $Z$ data can help to determine the
value of $g_2$, which is the coefficient of the $\ln(Q/2 Q_0)$
term in Eq.~(\ref{LY_form}). Given that, one can hope to study
the $x$ dependence of the non-perturbative function in more
detail. We also confirmed that it is reasonable to use a single
non-perturbative parameter $\overline{g_1}$ to fit  $Z$
boson data, and use that result to study the $Q_T$ distribution of
the $W^\pm$ boson for $Q_T<10$\,GeV. Recently this point has been made in
the context of a momentum-space fit \cite{qtres} using a single parameter.
Such an approach 
might indeed alleviate the computational overhead
required in order to do a complete Fourier transform in order to produce 
distributions of $W$ bosons and decay leptons
 necessary for $M_W$
analyses. However, if one is interested in testing the universality
property of the CSS resummation
formalism or making predictions about $W$ and $Z$ boson
production at future colliders, such as the CERN Large Hadron Collider,
then one must include the $Q$ (and, possibly, $x$) dependent term in the
non-perturbative function.

\section*{Acknowledgments}

We thank C.~Bal\'azs, D.~Casey, J.~Collins, W.-K.~Tung and the 
CTEQ collaboration for useful discussions.
This work  was supported in part
by National Science Foundation grants PHY-9514180 and PHY-9802564.
\newpage


\end{document}